# Pore-scale simulation of multicomponent multiphase reactive transport with dissolution and precipitation

Li Chen [a,b], Qinjun Kang [b*], Qing Tang [c], Bruce A. Robinson [d], Ya-Ling He [a], Wen-Quan Tao [a]

a: Key Laboratory of Thermo-Fluid Science and Engineering of MOE, School of Energy and Power Engineering, Xi'an Jiaotong University, Xi'an, Shaanxi 710049, China

b: Computational Earth Science Group (EES-16), Los Alamos National Laboratory, Los Alamos, New Mexico, USA

c: Shell Intl. Exploration & Production, Inc, 5521 Gasmer Dr., Houston, Texas, 77035, USA

d: Civilian Nuclear Programs, Los Alamos National Laboratory, Los Alamos, New Mexico 87545, USA

**Corresponding author:** Qinjun Kang

Email: qkang@lanl.gov

Phone:   (505) 665-9663

Fax:   (505) 665-8737

**Abstract**

Multicomponent multiphase reactive transport processes with dissolution-precipitation are widely encountered in energy and environment systems. A pore-scale two-phase multi-mixture model based on the lattice Boltzmann method (LBM) is developed for such complex transport processes, where each phase is considered as a mixture of miscible components in it. The liquid-gas fluid flow with large density ratio is simulated using the multicomponent multiphase pseudo-potential LB model; the transport of certain solute in the corresponding solvent is solved using the mass transport LB model; and the dynamic evolutions of the liquid-solid interface due to dissolution-precipitation are captured by an interface tracking scheme. The model developed can predict coupled multiple physicochemical processes including multiphase flow, multicomponent mass transport, homogeneous reactions in the bulk fluid and heterogeneous dissolution-precipitation reactions at the fluid-solid interface, and dynamic evolution of the solid matrix geometries at the pore-scale. The model is then applied to a physicochemical system encountered in shale gas/oil industry involving multiphase flow, multicomponent reactive transport and dissolution-precipitation, with several reactions whose rates can be several orders of magnitude different at a given temperature. The pore-scale phenomena and complex interaction between different sub-processes are investigated and discussed in detail.

# 1. Introduction

Multicomponent multiphase reactive transport processes in porous media are widespread in energy and environment sciences. Typical examples are proton exchange membrane fuel cells (PEMFC)[1], Liesegang phenomena [2], hydrocarbon recovery [3], geological storage of nuclear wastes [4], and carbon dioxide ($CO_2$) sequestration [5-7]. The coupled transport processes and multiphase phenomena, together with the dynamic evolutions of the solid matrix structures caused by dissolution-precipitation, make the situation very complicated [2, 4-8]. Usually the phenomena at the continuum-scale are of engineering interest; however they originate from the underlying pore-scale processes and interactions. For example, surface reactions taking place at the pore scale lead to the dissolution/precipitation of the solid phase, which in turn affect the fluid flow and solute transport, thus altering the material properties such as porosity and permeability that determine the continuum-scale behaviors. Therefore, understanding the meso/microscopic phenomena and the complicated interactions between different processes is critical to achieving a better prediction of the related systems.

With the improvement of the computational resources, pore-scale simulation, in which the realistic porous structures are considered, has become a critical technique for exploring multiple physicochemical problems. To accurately simulate pore-scale multicomponent multiphase reactive transport processes with dynamic evolutions of the solid phases, where the topography of the evolving liquid-gas-solid interfaces is part of the required result, the numerical model must be able to handle the following four major fundamental issues. The first one is to capture the deformable liquid-gas interfaces which may stretch, break-up or coalesce. Available models for tracking the moving liquid-gas interface can be roughly classified in two groups: the diffusion interface model and the sharp interface model and one can refer to a recent review paper for more details [9]. The second fundamental issue is to track the moving fluid-solid interfaces due to dissolution-precipitation (or melting-solidification). The common feature of the problems with the evolutions of solid phases is that there is no fluid flow and mass transport in the solid phase, and flow at the solid-fluid interfaces is usually subjected to a no-slip condition as the dissolution-precipitation (or melting-solidification) is relatively slow in the practical systems [4-6, 8]. The third one is to simulate mass transfer in the multiphase systems. The mass transfer processes in different phases, which are very important for understanding the reactive transport processes, pose a great challenge for simulation. In addition, the accompanying evolution of the liquid-gas-solid interfaces makes the inter-phase mass transfer more complicated [4, 9]. The last one is the incorporation of homogeneous and heterogeneous reactions into the simulations [10].

Different numerical methods have been developed to investigate pore-scale fluid flow and transport processes, including the direct numerical simulation method (DNS)[6], pore-network model (PNM)[11], lattice Boltzmann method (LBM) [4, 12-18], and smoothed particle hydrodynamics [19, 20]. Each of these methods has its own advantage and disadvantages. Little work has been done on the development of pore-scale reactive transport models with dissolution-precipitation because of the difficulty of handling the coupled physicochemical processes and directly capturing the complex structural evolutions. Single-phase fluid flow and reactive transport with dissolution-precipitation was studied by Kang et al. [12, 13] in which the LBM was used to simulate transport processes and the Volume of Pixel (VOP, a method for tracking fluid-solid interfaces) was adopted to update the solid structures. They predicted the relationship between permeability and porosity under different $Pe$ (Peclet number, representing the relative strength of convection to diffusion) and $Da$ (Damköhler number, representing the relative strength of reaction to diffusion). Later, Kang et al. [10] extended their model to multicomponent systems and used the model to study reactive transport processes associated with geological $CO_2$ sequestration [5]. The numerical studies of multiphase fluid flow coupled with reactive transport with moving solid-fluid interfaces are scarce in the literature [4, 15]. Recently, Parmigiani et al. [15] used the LBM to study the injection process of a non-wetting fluid into a wetting fluid coupled with dynamic evolution of the solid geometries. In their study the change of solid phase was caused by melting only. Very recently, Chen et al. [4] constructed a pore-scale model based on the LBM and the VOP to simulate multiphase reactive transport with phase transition and dissolution-precipitation processes [4]. Their pore-scale model can capture coupled non-linear multiple physicochemical processes including multiphase flow with phase separation, mass transport, chemical reaction, dissolution-precipitation, and dynamic evolution of the pore geometries. The model was used to study the thermal migration of a brine inclusion in a salt crystal [4]. However, in their model, the two-phase system is a single-component water-vapor system and only the transport of a single solute was considered [4]. From the above review, it can be seen that currently there are no pore-scale studies of multicomponent multiphase reactive transport processes with dissolution-precipitation.

The objective of the present study is to develop a pore-scale model which can handle multiphase flow involving multicomponent homogenous and heterogeneous reactions. Specific goals include (i) constructing a two-phase multi-mixture model based on the LBM, (ii) simulating multiple physicochemical processes including multiphase fluid flow, multicomponent mass transport, homogeneous reactions in the bulk fluid and heterogeneous reactions at the fluid-solid interfaces, and dissolution and precipitation of the solid phases at the pore-scale, and (iii) investigating the complex interplay between different sub-processes. We base our model on the LBM, which is well suited for solving fluid flow in complex geometries and has been successfully used in the study of flow in porous media [21]. Furthermore, the kinetic nature of the LBM enables it to conveniently represent microscopic interactions between different fluids, thereby facilitating the automatic tracking of the fluid-fluid interfaces in a multiphase system [22-24]. The fluid-solid interactions can also be implemented conveniently in the LBM without including additional complex kernels [25, 26]. To the best of our knowledge, there has been no such a pore-scale model reported in the literature. The specific physicochemical problem considered in this study stems from a novel unconventional oil recovery process involving multicomponent multiphase reactive transport with dissolution and precipitations. However, readers with background in PEMFC, micro-reactors, oil recovery, and geological storage of $CO_2$ and nuclear wastes may find potential applications of this model.

The rest of this paper is organized as follows. In Section 2 the physicochemical problem considered is briefly introduced. In Section 3 the multicomponent multiphase pseudopotential LB model proposed by Shan and Chen and the mass transport LB model are introduced. Then a two-phase multi-mixture model, in which the pseudopotential LB model, the mass transport LB model and a front-tracking method for liquid-solid interface are incorporated, is developed in Section 4. In Section 5, the two-phase multi-mixture model is adopted to simulate the physicochemical problem introduced in Section 2 and investigate the complex interactions between different processes. Finally some conclusions are presented in Section 6.

## 2. Physicochemical problem

The physicochemical system considered in this study is encountered during the development of a novel unconventional oil recovery technology, whose viability relies heavily on a better understanding of the pore-scale processes. The reactions involved are as follows

$$A_{(aq)} \rightleftarrows B_{(aq)} + 0.5C_{(g)} \tag{1}$$

$$B_{(aq)} + D_{(s)} \rightarrow E_{(aq)} \tag{2}$$

$$B_{(aq)} + E_{(aq)} \rightarrow F_{(aq)} \tag{3}$$

$$F_{(aq)} \rightarrow F_{(s)} \tag{4}$$

Reaction R1 and R3 are homogeneous reactions in the bulk fluid, while the remaining two are heterogeneous reactions at the liquid-solid interface. The physicochemical processes are generalized as follows (shown in Fig. 1). Initially, the system contains an aqueous component $A_{(aq)}$, a gas component $C_{(g)}$ and a primary solid phase $D_{(s)}$. Then due to chemical disequilibrium, $A_{(aq)}$ is decomposed into an aqueous component $B_{(aq)}$ and the gas species $C_{(g)}$ (Reaction 1), both in the bulk liquid and at the surface of $D_{(s)}$, but the reaction rate at the surface is 1000 times that in the bulk. $B_{(aq)}$ will react with the primary solid phase $D_{(s)}$ (Reaction 2), which leads to the dissolution of $D_{(s)}$ and the production of $E_{(aq)}$. $E_{(aq)}$ will react with $B_{(aq)}$ (Reaction 3), generating an aqueous component $F_{(aq)}$. Under certain circumstances, $F_{(aq)}$ will precipitate and a secondary solid phase $F_{(s)}$ is generated on the surface of $D_{(s)}$ (Reaction 4). The reaction kinetics of the above reactions are as follows:

$$R_1 = k_1(C_{A_{(aq)}} - C_{B_{(aq)}}(C_{C_{(g)}})^{0.5}/K_{eq}) \tag{5a}$$

$$R_2 = k_2 C_{B_{(aq)}} \tag{5b}$$

$$R_3 = k_3 C_{B_{(aq)}} C_{E_{(aq)}} \tag{5c}$$

$$R_4 = k_4(C_{F_{(aq)}} - C_{F_{(aq)},sa}) \tag{5d}$$

where $C$ is the concentration and subscript "sa" denotes saturation concentration. The equilibrium constant $K_{eq}$ for Reaction 1 and reaction rate constant $k$ of the four reactions are listed in Table 1.

**Table 1** Parameters used in the simulations

The sub-processes involved in the above physicochemical problem include multiphase fluid flow, multicomponent transports, homogeneous and heterogeneous reactions, and dissolution and precipitations of the solid phases. The sub-processes and interactions between them are described by a set of non-linear partial differential equations, including Navier-Stokes equation for multiphase fluid flow, convection-diffusion equations for temperature and concentrations of multiple components, and the volume evolution equations for the solid phase, combined with proper initial and boundary conditions. The complex processes and complicated interactions involved pose great challenges to the numerical simulations. Compared to conventional finite-element, finite-volume and finite-difference technique, the LBM is more promising in dealing with the complicated non-linear characteristics as well as the complex structures. In this study, we propose a two-phase multi-mixture model based on the LBM to solve the above processes. Before establishing this model, the pseudopotential LB model for multicomponent multiphase flow [22, 23] and the mass transport LB model for species transport [27, 28], which are the cornerstones of the two-phase multi-mixture model, are introduced in the Section 3.

**Fig. 1** The schematic of the multicomponent multiphase reactive transport processes which include liquid-gas phases, five components, two solid phases and four reactions

## 3. Lattice Boltzmann models

*3.1. The pseudopotential LB model for multicomponent multiphase flow*

In the LB method, the motion of fluid is described by a set of particle distribution functions. Based on the simple and popular Bhatnagar-Gross-Krook (BGK) collision operator, the evolution equation of the distribution function is written as

$$f_{\sigma,\alpha}(\mathbf{x}+c\mathbf{e}_\alpha \Delta t, t+\Delta t) - f_{\sigma,\alpha}(\mathbf{x},t) = -\frac{1}{\tau_{\sigma,\upsilon}}(f_{\sigma,\alpha}(\mathbf{x},t) - f_{\sigma,\alpha}^{eq}(\mathbf{x},t)) + \omega_\alpha S_\sigma \qquad (6)$$

where $f_{\sigma,\alpha}(x,t)$ is the density distribution function of $\sigma$ component at the lattice site **x** and time $t$, $f^{eq}$ is the equilibrium distribution function, $c=\Delta x/\Delta t$ is the lattice speed with $\Delta x$ and $\Delta t$ as the lattice spacing and time step, respectively, and $\tau_v$ is the dimensionless relaxation time. $\omega_\alpha$ are weight factors and will be introduced in Eq. (8). $S_\sigma$ is the source term related to homogeneous reaction and will be defined in Section 4. For D2Q9 model with nine velocity directions at a given point in two-dimensional space, discrete velocities $\mathbf{e}_\alpha$ are given by

$$\mathbf{e}_\alpha = \begin{cases} 0 & \alpha = 0 \\ (\cos[\frac{(\alpha-1)\pi}{2}], \sin[\frac{(\alpha-1)\pi}{2}]) & \alpha = 1,2,3,4 \\ \sqrt{2}(\cos[\frac{(\alpha-5)\pi}{2}+\frac{\pi}{4}], \sin[\frac{(\alpha-5)\pi}{2}+\frac{\pi}{4}]) & \alpha = 5,6,7,8 \end{cases} \quad (7)$$

The equilibrium distribution functions $f^{eq}$ for the D2Q9 lattice are of the form

$$f_{\sigma,\alpha}^{eq} = \omega_\alpha \rho_\sigma \left[1 + \frac{3}{c^2}(\mathbf{e}_\alpha \cdot \mathbf{u}) + \frac{9}{2c^4}(\mathbf{e}_\alpha \cdot \mathbf{u})^2 - \frac{3}{2c^2}\mathbf{u}^2\right] \quad (8)$$

The weight factors $\omega_\alpha$ are given by $\omega_0=4/9$, $\omega_{1-4}=1/9$, and $\omega_{5-8}=1/36$. The fluid density $\rho_\sigma$ and velocity $\mathbf{u}_\sigma$ can be obtained from the first and second moments of the density distribution functions

$$\rho_\sigma = \sum_\alpha f_{\sigma,\alpha} \quad \text{(9-a)}$$

$$\rho_\sigma \mathbf{u}_\sigma = \sum_\alpha f_{\sigma,\alpha} \mathbf{e}_{\sigma,\alpha} \quad \text{(9-b)}$$

The viscosity in the lattice unit is related to the collision time by

$$\upsilon_\sigma = c_s^2(\tau_{\sigma,\upsilon} - 0.5)\Delta t \quad (10)$$

where $c_s$ is the lattice sound speed.

Microscopically, the segregation of a fluid system into different phases results from the forces between molecules. For a multicomponent system, there exist both the intra-molecule interaction force and inter-molecule interaction force. In Shan and Chen's pseudopotential model [22, 23] an effective mass $\psi$ based on local fluid density is introduced to calculate the interaction force. The interaction force between the particles of the same component can be written as

$$\mathbf{F}_{\sigma\sigma} = -G_{\sigma\sigma}\psi_\sigma(\mathbf{x})\sum_{\alpha=1}^{N} w(|\mathbf{e}_\alpha|^2)\psi_\sigma(\mathbf{x}+\mathbf{e}_\alpha)\mathbf{e}_\alpha \qquad (11)$$

where $G_{\sigma\sigma}$ reflects the interaction strength, and $w(|\mathbf{e}^\alpha|^2)$ are the weights. If only the interactions of four nearest neighbors with $|\mathbf{e}^\alpha|^2 =1$ and the four next-nearest neighbors $|\mathbf{e}^\alpha|^2 =2$ are considered, $w(1)=1/3$ and $w(2)=1/12$. For higher order isotropy which can reduce the spurious currents, one can refer to [29, 30]. In the present study, two components are considered with $\sigma = 1$ and 2 as the liquid component and gas component, respectively. The gas component is considered as an ideal fluid and thus $G_{22}$ is zero. The liquid component, on the other hand, is considered as a non-ideal fluid with Carnahan-Starling (C-S) equation of state (EOS) [31]

$$p = \rho RT \frac{1+b\rho/4+(b\rho/4)^2-(b\rho/4)^3}{(1-b\rho/4)^3} - a\rho^2 \qquad (12)$$

where $a = 0.4963(RT_c)^2/p_c$, $b = 0.1873RT_c/p_c$. Thus, effective mass $\psi_1$ of the liquid component is expressed as [31]

$$\psi_1 = \sqrt{\frac{6}{G_{11}}(\rho_1 RT \frac{1+b\rho_1/4+(b\rho_1/4)^2-(b\rho_1/4)^3}{(1-b\rho_1/4)^3} - a\rho_1^2 - c_s^2\rho_1)} \qquad (13)$$

Other $\psi_1$ can be used based on different EOSs. Note that $G_{11}$ is cancelled out in the EOS and is only required to guarantee the whole term inside the square root of Eq. (13) to be positive. In the current study, $G_{11}$ is set to $-1$. In the simulation of Yuan and Schaefer [31], $a = 1$, $b = 4$ and $R =1$, $T_c = 0.094$ and $\rho_c=0.13044$. From preliminary simulations we find that the simulation of the multicomponent multiphase system tends to diverge when the vapor density is too small. This is because when the vapor density is too small, the vapor velocity will change significantly even a small force is added. Because large spurious currents are believed to cause the numerical instability [29, 31, 32], hence a small vapor density is not desirable. Therefore, in the present study we set $a = 0.09926$, $b = 0.18727$, $R = 0.2$, $T_c = 1.0$ and $\rho_c=5.469129$ to increase the vapor density while still maintaining the desired density ratio of liquid to vapor. For example, when $T=0.7T_c$, the liquid and vapor densities are 7.739 and 0.17, respectively, greatly increased compared with the original one as 0.3589 and 0.00670, respectively, while the density ratio of 45.5 is not much different from the original one of 53.6. It is worth mentioning that recently there have been a few studies devoted to two-phase flow with large density and viscosity ratio using Shan and Chen pseudopotential model [33-35]. For a recent review of the theory and application of the pseudopotential model one can refer to [24].

An expression similar to Eq. (11) can be used to calculate the inter-molecule interaction force [36], with a proper choice of effective mass for the gas component, such as $\psi_2=\rho_2$ in [33] and $\psi_2=\sqrt{\rho_2}$ in [37]. However, we find from preliminary simulations that the inter-molecule interaction force calculated this way results in a mutual diffusivity which is too small to let the system to reach equilibrium even after long time simulations, and leads to non-uniform distributions of the components in the gas phase. To fix this problem, we use the following expressions to calculate the inter-molecule interaction force which is attractive in the gas phase to enhance the mutual diffusivity [38]

$$F_{12} = -G_{12}\varphi_1(\mathbf{x})\sum_{\alpha=1}^{N} w(|\mathbf{e}_\alpha|^2)\varphi_2(\mathbf{x}+\mathbf{c}_\alpha)\mathbf{c}_\alpha \tag{14-a}$$

$$F_{21} = -G_{21}\varphi_2(\mathbf{x})\sum_{\alpha=1}^{N} w(|\mathbf{e}_\alpha|^2)\varphi_1(\mathbf{x}+\mathbf{c}_\alpha)\mathbf{c}_\alpha \tag{14-b}$$

where $\varphi_1$ and $\varphi_2$ is different from $\psi_1$ and $\psi_2$ and is designed as follows [38]

$$\varphi_1(\rho_2) = 1 - \exp(-\rho_2/\rho_{20}) \tag{15-a}$$

$$\varphi_2(\rho_1) = a_0 - \exp(-\rho_1/\rho_{10}) \tag{15-b}$$

The choice of $G_{12}$, $a_0$, $\rho_{10}$ and $\rho_{20}$ is critical for the multicomponent multiphase system and is tedious work. From lots of preliminary simulations we find that generally higher $G_{12}$ and $|\rho_{10}|$ lead to higher mutual diffusivity in the gas phase. In the present study, we set $a_0=0.005$, $\rho_{10}=-0.008/\log(a_0)$, $\rho_{20}=0.003$ and $G_{12}=G_{21}=0.001$.

Besides the fluid-fluid interaction, there are also forces between fluid and solid phase, which is calculated by

$$\mathbf{F}_{\sigma w} = -w_{\sigma w}\psi_\sigma(\mathbf{x})\sum_{\alpha=1}^{N} w(|\mathbf{e}_\alpha|^2)s(\mathbf{x}+\mathbf{e}_\alpha)\mathbf{e}_\alpha \tag{16}$$

where *w* determines the strength of the interaction between fluid and the solid wall, by adjusting which different contact angles can be obtained. *s* represents the wall density, with a value of 1 for solid nodes and 0 for fluid nodes.

The correct incorporation of the forcing terms into the LBM is an important issue. Recently, different force schemes are assessed [39-41]. In the current study, the Shan and Chen original force scheme is adopted [23] where the forces acting on the $\sigma$ component are accounted for by modifying the velocity in Eq. (8) as

$$\mathbf{u} = \mathbf{u}' + \frac{\tau_\sigma \mathbf{F}_\sigma}{\rho_\sigma} \qquad (17)$$

$\mathbf{F}_\sigma$ is the total forces acting on the $\sigma$ component. $\mathbf{u}'$ is the common velocity of all the components defined as

$$\mathbf{u}' = \frac{\sum_\sigma \rho_\sigma \mathbf{u}_\sigma / \tau_\sigma}{\sum_\sigma \rho_\sigma / \tau_\sigma} \qquad (18)$$

Finally, the actual physical velocity is the average velocity before and after the collision and is given by

$$\mathbf{u}_\mathrm{r} = \frac{\sum_\sigma \rho_\sigma \mathbf{u}_\sigma + \frac{\Delta t}{2} \sum_\sigma \mathbf{F}_\sigma}{\sum_\sigma \rho_\sigma} \qquad (19)$$

Validation of our codes for the multicomponent multiphase pseudopotential model with large density ratio can be found in Appendix A.

*3.2 LB model for aqueous species transport*

The following evolution of the LB equation is used to describe the aqueous species transport

$$g_{k,\alpha}(\mathbf{x} + c\mathbf{e}_i \Delta t, t + \Delta t) - g_{k,\alpha}(\mathbf{x}, t) = -\frac{1}{\tau_{k,g}}(g_{k,\alpha}(\mathbf{x}, t) - g_{k,\alpha}^{eq}(\mathbf{x}, t)) + J_{k,\alpha} \Delta t S_k \qquad (20)$$

where $g_{k,\alpha}$ is the concentration distribution function for the *k*th species and $S_k$ is the source term related to reaction. $J_{k,\alpha}$ is the rest fraction and will be given in Eq. (22).

Note that the convection-diffusion equation is linear in velocity **u**. This means that the equilibrium distributions for scalar transport need only to be linear in **u**; and thus lattices with fewer vectors can be used for scalar transport. A reduced D2Q5 lattice model of the original D2Q9 model is used, in which the lattice vectors in the diagonal directions ($e_\alpha$, $\alpha$=5-8) in Eq. (7)) are not used. An equilibrium distribution that is linear in **u** is adopted [28]

$$g_{k,\alpha}^{eq} = C_k \left[ J_{k,\alpha} + \frac{1}{2} \mathbf{e}_\alpha \cdot \mathbf{u} \right] \quad (21)$$

where $C_k$ is the solute concentration. $J_{k,\alpha}$ is given by [28]

$$J_{k,\alpha} = \begin{cases} J_{k,0}, & \alpha = 0 \\ (1-J_{k,0})/4, & \alpha = 1,2,3,4 \end{cases} \quad (22)$$

where the rest fraction $J_0$ can be selected from 0 to 1. The equilibrium distribution function given by Eq. (21) can cover a wide range of diffusivity by adjusting $J_0$, which is a prominent advantage of such an equilibrium distribution. Note that different forms of equilibrium distribution functions with D2Q5 are employed in literature. For example, in the study of Huber et al. [42], the equilibrium distribution function is chosen with $J_0$=1/3 and $J_{1-4}$=1/6, and in the study of Kang et al. [43] with $J_0$=0 and $J_{1-4}$=1/4. Eq. (21) is a general formula. The accuracy and efficiency of the reduced D2Q5 model has been confirmed in many studies [4, 36, 43, 44].

The concentration is obtained by

$$C_k = \sum g_{k,\alpha} \quad (23)$$

The diffusivity is related to the relaxation time by

$$D_k = \frac{1}{2}(1-J_{k,0})(\tau_{k,g} - 0.5) \quad (24)$$

Validations of the LB mass transport model can be found in Appendix B.

## 4. Establishment of a two-phase multi-mixture model

The physicochemical problem studied contains five components, one of which is a gas component ($C_{(g)}$) and the remaining four ($A_{(aq)}$, $B_{(aq)}$, $E_{(aq)}$ and $F_{(aq)}$) are aqueous components which are miscible with each other in the liquid phase. Modeling such a system with so many components is a great challenge. Theoretically a multicomponent system can be simulated using the original multicomponent multiphase Shan-Chen model [45, 46], with each component corresponding to a distribution function. Recently, Kamali et al. [37] used such a scheme to simulate the reactive transport process in a 1D Fischer-Tropsch reactor involving a aqueous component and three gas components. Such an approach, however, requires considerable preliminary work for tuning the inter-molecular interaction strength $G_{ij}$ and corresponding parameters of $a_0$, $\rho_{10}$ and $\rho_{20}$ for every pair of components. For a system containing $n$ components, it means that $4 C_n^2$ (C denotes combination function and $C_n^2$ means the number of combinations by randomly choosing two from $n$.) parameters should be determined. As an example, relevant to the system of the present study, five components are included and thus 40 parameters are required to be carefully tuned, which is really a tedious job. Besides, for a multicomponent multiphase system which contains several aqueous components as in the present study, a stable system is very difficult to achieve, compared to the system considered in [37], which contains only one aqueous component with the remaining components as gaseous components. This is because for the former case, the strong interactions between aqueous components can render the simulation unstable. However, for the latter case, once a single-component two-phase system is established, other gaseous components can be easily added into the system thanks to the relatively weak interactions between the aqueous component and the gas components (set as 0.001 in [37]).

Thus in the present study, alternatively, a two-phase multi-mixture model is proposed to simulate the physicochemical reactive transport processes involving multiple components. In this model, each phase is considered as a mixture of all components in this phase. For example, for a system contains $n$ gas components and $m$ aqueous components, the gas phase and the liquid phase are considered to be the mixture of the $n$ gas components and the $m$ aqueous components, respectively. The properties of each phase can be calculated from the properties of the corresponding components involved, using the linear interpolation based on the mass fraction of each component. For example, the density of gas phase and liquid phase can be calculated by

$$\rho_g = \sum_{k=1}^{n} C_k M_k \tag{25-a}$$

$$\rho_l = \sum_{k=1}^{m} C_k M_k \qquad (25\text{-}b)$$

respectively, where $M_k$ is the molar mass of $k$th component. The other properties such as viscosity can be calculated in the similar way.

### 4. 1. Multicomponent mass transport

After the two-phase system is constructed using the scheme in Section 3.1, the transport of each component is then solved based on the two-phase flow field. To properly account for the partitioning of $C_k$ between the phases in computational cells containing the interface, it is convenient to write the following two transport equations for the concentration $C_{k,\text{liquid}}$ and $C_{k,\text{gas}}$ in each phase [47]

$$\frac{\partial C_{k,\text{gas}}}{\partial t} + (\mathbf{u} \cdot \nabla) C_{k,\text{gas}} = D_{k,\text{gas}} \Delta C_{k,\text{gas}} + S_{k,\text{gas}} + S_{\text{interface}} \qquad (26\text{-}a)$$

$$\frac{\partial C_{k,\text{liquid}}}{\partial t} + (\mathbf{u} \cdot \nabla) C_{k,\text{liquid}} = D_{k,\text{liquid}} \Delta C_{k,\text{liquid}} + S_{k,\text{liquid}} + S_{\text{interface}} \qquad (26\text{-}b)$$

The source term $S_{\text{interface}}$ denotes mass transfer across the two-phase interface and is only non-zero in the computational cells containing phase interface. In this study, the aqueous solutes transport only in the liquid phase. Therefore, only Eq. (26-b) is solved and the liquid-gas interface is considered as a non-flux boundary for the transport of aqueous solutes which leads to $S_{\text{interface}} = 0$ in Eq. (26-b). Similarly, for a gas component that mainly transports in the gas phase and its transport in liquid phase can be neglected, only Eq. (26-a) is solved and the liquid-gas interface is also considered as a non-flux boundary for the gas component with $S_{\text{interface}}$ in Eq. (26-a) as zero. For the case where a component transports in both phases and the across-interface phenomena are important, one can refer to Ref. [48] for more information. Note that in gas (liquid) phase, only $n$-1 ($m$-1) transport equations need to be solved, and the concentration of the remaining gas (aqueous) component is obtained by subtracting the sum of the concentrations of the $n$-1 ($m$-1) components from the total gas (liquid) concentration.

In the present study the gas component $C_{(g)}$ is generated in the liquid phase by Reaction R1, and it will transport from the liquid phase, across the liquid-gas interface and finally into the gas phase. Therefore the transport of $C_{(g)}$ should be solved in the entire two-phase domain. This is achieved by solving the evolution equation of the gas component distribution function (Eq. (6) with σ=2) [45, 46], as $C_{(g)}$ is the only gas component in the present study. The $C_{(g)}$ generated in the liquid phase is added into the evolution equation as a source term.

In summary, when using the two-phase multi-mixture model to simulate the physicochemical processes considered in the present study, firstly we use the Shan-Chen model to construct the two-phase system. Here the gas phase consists of only $C_{(g)}$ and is described by $f_2$ in the Shan-Chen model, while the liquid phase is a mixture of $A_{(aq)}$, $B_{(aq)}$, $E_{(aq)}$ and $F_{(aq)}$ and is described by $f_1$ in the Shan-Chen model. Then we solve the transport equation for $B_{(aq)}$, $E_{(aq)}$ and $F_{(aq)}$ in the liquid phase with the liquid-gas interface as a non-flux boundary. The concentration of $A_{(aq)}$ is obtained by subtracting the sum of the concentrations of $B_{(aq)}$, $E_{(aq)}$ and $F_{(aq)}$ from the total liquid concentration. Solving the transport of $C_{(g)}$ independently using LB mass transport model is not required as it has already been considered by the evolution of $f_2$ in the Shan-Chen model, as discussed previously. Note that such a two-phase multi-mixture model has been successfully used for simulating multicomponent multiphase reactive transport in PEMFC. In [49, 50], the liquid phase only contains liquid water while the gas phase is a mixture of oxygen, nitrogen and water vapor. The numerical method in [49, 50] is based on conventional FVM techniques.

It is worth mentioning that, in the Shan-Chen multicomponent multiphase model, the diffusivity of each component depends on its collision time as well as its interactions with other components [45, 46]. Thus, the diffusivity of $C_{(g)}$ can be adjusted to coincide with the physical one by tuning the collision time of that component [37].

*4. 2. Homogeneous reactions in the bulk fluid*

The homogeneous reactions are considered in the evolution equations of distribution functions. For reaction R1, the production of $C_{(g)}$ contributes to the loss of mass of the liquid phase and the gain of mass of the gas phase. Thus, the source term $S_\sigma$ in Eq. (6) is

$$S_1 = -0.5 R_1 M_{C_{(aq)}}, \ S_2 = 0.5 R_1 M_{C_{(aq)}} \tag{27}$$

where $M$ is the molar mass. The transport of aqueous components $B_{(aq)}$, $E_{(aq)}$ and $F_{(aq)}$ is handled by Eq. (20). Their source terms related to homogeneous reactions are as follows

$$S_{B_{(aq)}} = R_1 - R_3, \ S_{E_{(aq)}} = -R_3, \ S_{F_{(aq)}} = R_3 \qquad (28)$$

*4. 3. Heterogeneous reactions at the solid-liquid interface*

At the liquid-$D_{(s)}$ interfaces, heterogeneous dissolution reaction R2 takes place, which consumes $B_{(aq)}$ and produces $E_{(aq)}$. The consumption of $B_{(aq)}$ and the production of $E_{(aq)}$ are taken into account as the boundary conditions for the transport of $B_{(aq)}$ and $E_{(aq)}$

$$D_{B_{(aq)}} \frac{\partial C_{B_{(aq)}}}{\partial n} = -k_2 C_{B_{(aq)}} \qquad (29)$$

$$D_{E_{(aq)}} \frac{\partial C_{E_{(aq)}}}{\partial n} = k_2 C_{B_{(aq)}} \qquad (30)$$

where *n* denotes direction normal to the solid surface pointing to the void space.

The initially existing $D_{(s)}$ can be regarded as the impurity for the precipitation of $F_{(aq)}$, and thus the precipitation occurs as crystal growth on the solid $D_{(s)}$ surfaces, with homogeneous nucleation excluded [2]. Once $F_{(s)}$ appears in the system, its surface is also allowed for crystal growth. The boundary condition for the precipitation reaction at the liquid-$D_{(s)}$ (or $F_{(s)}$) interface is

$$D_{F_{(aq)}} \frac{\partial C_{F_{(aq)}}}{\partial n} = k_4 (C_{F_{(aq)}} - C_{F_{(aq)},sa}) \qquad (31)$$

The implementation of the above boundary conditions for dissolution and precipitation reactions in the framework of LBM will be discussed in Section 4.6. The validation of the dissolution-precipitation model can be found in Appendix C.

*4.4. VOP method for update of solid phases*

Due to the dissolution reaction, the volume of $D_{(s)}$ is reduced according to the following equation

$$\frac{dV_{D_{(s)}}}{dt} = -A \overline{V_{D_{(s)}}} k_2 C_{B_{(aq)}} \qquad (32)$$

where $V_D$, $\overline{V_D}$ and $A$ are the dimensionless volume, molar volume and specific surface area of $D_{(s)}$, respectively. Thus, $V_D$ is updated at each time step explicitly according to the following equation

$$V_{D_{(s)}}(t+\Delta t) = V_{D_{(s)}}(t) - A\overline{V_{D_{(s)}}}k_2 C_{B_{(aq)}} \Delta t \tag{33}$$

where $\Delta t$ is the time step and equals that in Eq. (8) in the simulation.

Similarly, the volume of $F_{(s)}$ is increased by

$$\frac{dV_{F_{(s)}}}{dt} = -A\overline{V_{F_{(s)}}}k_4 C_{F_{(aq)}} \tag{34}$$

and is updated at each time step according to

$$V_{F_{(s)}}(t+\Delta t) = V_{F_{(s)}}(t) - A\overline{V_{F_{(s)}}}k_4 C_{F_{(aq)}} \Delta t \tag{35}$$

We use the VOP method developed by Kang et al. [10] to track and update the liquid-solid interfaces. For more information of the VOP method, one can refer to [10, 12].

### 4.5. Special treatment of nodes undergoing phase change

The liquid-gas interfaces serve as non-flux boundaries for the transport of $A_{(aq)}$, $B_{(aq)}$, $E_{(aq)}$ and $F_{(aq)}$. We use the following simple approach to distinguish the liquid phase and the gas phase: if the density of a node is higher than a critical density (chosen as a half of the maximum density in the system), the node is regarded as a liquid node, otherwise it is a gas node. Thus the liquid-gas interface is connected by marked nodes. It is well known that the spurious currents around the interface increase with the liquid-gas density [31, 32]. Such large spurious currents have unphysical effects on the mass transport, and thus must reduced [32]. Close examination reveals that large spurious currents around the liquid-gas phase are mainly in the gas phase [4], and thus they are avoided in the present study as we only solve the mass transport in the liquid phase.

Since the liquid-gas interfaces are deformable, the status of a computational node can change from liquid to gas, or vice versa (shown in Fig. 2). Besides, the switch of status between liquid-solid phases for a computational node also takes place due to dissolution-precipitation (also shown in Fig. 2). Such a feature poses significant challenges to the pore-scale simulation. Special treatment of nodes undergone status change (liquid-gas or liquid-solid) is required to guarantee the conservation of mass, momentum and concentration. The scheme proposed in our recent paper [4] is employed.

**Fig. 2** Dynamic evolutions of liquid-gas-solid interfaces during the multicomponent multiphase reactive transport processes. There are three types of nodes in the domain: solid nodes, liquid nodes, and gas nodes; and there are four kinds of changes of node type: namely liquid to solid due to precipitation, solid to liquid due to dissolution, liquid to gas, and gas to liquid. The last two of these changes of node type are due to evolution of the liquid-gas interfaces. Note that there is no exchange between gas node and solid node as dissolution and precipitation only take place at liquid-solid interface

*4.6. Moving and reactive boundary*

In this study, the liquid-gas interface is a moving boundary and the liquid-solid interface is a reactive boundary. A general concentration boundary condition in the framework of LBM is required to handle reactive and moving boundaries with complex structures. Several concentration boundary conditions have been developed for different boundary types in the literature [43, 51]. However, the work specifically devoted to reactive and moving boundaries is limited. In Ref. [4], we proposed a new general LB concentration boundary condition for handling various concentration boundary situations based on the work of Kang et al. [43] and Zhang et al.[51]. This general boundary condition is employed in the present study.

*4.7. Numerical procedures*

In summary, the numerical procedure of the two-phase multi-mixture model for multicomponent multiphase reactive transport with dissolution-precipitation processes after the initialization step is as follows

Step 1: updating the flow field using the multicomponent multiphase pseudopotential model;

Step 2: if the liquid-gas interface is changed, the concentration fields for $A_{(aq)}$, $B_{(aq)}$, $E_{(aq)}$ and $F_{(aq)}$ are amended using the scheme outlined in Section 4.5, otherwise this step is skipped;

Step 3: based on the fluid flow filed obtained in Step 1, solving aqueous component transport in the liquid phase with homogeneous reactions in the bulk fluid and heterogeneous reactions at the liquid-solid interface using the LB mass transport model;

Step 4: updating the volume of solid nodes participating in the dissolution-precipitation reactions and then updating the geometry of the solid phase using the VOP method introduced in Section 4.4;

Step 5: if the liquid-solid interface is changed, the density, velocity and concentration fields are amended using the scheme outlined in Section 4.5;

Step 6: repeating Steps 1-5 until all the reactions in the system stop.

Our numerical model is constructed incrementally by combining different models including the multicomponent multiphase flow SC model [22, 23], mass transport with homogeneous and heterogeneous reactions [2, 27, 28, 36] and the VOP for dissolution and precipitation [2, 12-14]. Since no analytical solutions or experimental results for such complex physicochemical processes are available for quantitative comparisons with our simulation results, the validation of our model is achieved by testing each sub-model using certain physicochemical problems that have analytical solutions. Some of the validation tests have been presented in our previous paper [4, 36, 43, 44]. Additional validations are presented in the Appendix.

## 5. Results and discussion

The computational domain in the current study can be considered as a beaker with a solid particle of $D_{(s)}$ and the remaining space occupied by liquid and gas phases as shown in the inset image of Fig. 3. Since the present study is two-dimensional, the solid $D_{(s)}$ can be considered as a cylinder. The solid $D_{(s)}$ is completely immersed in the liquid phase, and its initial shape is circular for simplicity. The physical size of the computational domain is $0.1 \times 0.16$ m. The $D_{(s)}$ particle has a radius of 0.015 m and is located at (0.05m, 0.05m). The left, right and bottom boundaries are solid walls of the beaker, which are static for fluid flow and no-reactive for mass transport. The contact angles of the three walls are $90^0$. The top boundary is a free outlet with gradient of all the variables set to zero. The solid surfaces of $D_{(s)}$ and $F_{(s)}$ are static for fluid flow and their contact angles are set as $90^o$. Initially, the liquid phase only contains $A_{(aq)}$ and the reactions R1-R4 are turned on step by step depending on the emphasis of the investigation. The computational domain is discretized using $100 \times 160$ lattices, allowing for a resolution of $1 \times 10^{-3}$ m one lattice. The kinematic viscosities of both liquid mixture and gas phase are $\upsilon_\sigma = 1 \times 10^{-6}$ m$^2$ $s^{-1}$ with corresponding relaxation time $\tau_\sigma = 1.0$; and the diffusion coefficient of all the components in liquid phase is $D = 4.5 \times 10^{-8}\,m^2\,s^{-1}$ with corresponding $J_0$ in Eq. (22) and $\tau$ as 0.9 and 0.65, respectively. Setting the relaxation times and $J_0$ this way leads to equal time in each time step for fluid flow and mass transport, thus inner iteration for either process is not required. Dissolution reaction takes place on the surface of $D_{(s)}$ and precipitation reaction occurs on either the $D_{(s)}$ or the $F_{(s)}$ surface. The general concentration LB boundary condition for mass transport mentioned in Section 4.6 is employed on the reactive liquid-solid boundaries and the moving liquid-gas boundaries. The parameters are listed in Table 1.

*5.1 Tthe two-phase system*

We first establish the system using the Shan-Chen model introduced in Section 3.1, without considering any reactions. The initial density distribution of liquid component is $\rho_1$=8.31 and $\rho_1$=0.05 for $y$<0.1m and $y$>0.1m, respectively, while that for gas component is $\rho_2$=0.0001 and $\rho_2$=0.05 for $y$<0.1m and $y$>0.1m, respectively. The temperature is $T$=0.65$T_c$. Fig. 3 shows the density along the $y$ axis at $t = 50000$ step, which is the averaged density of the densities at $x$=0.02 and $x$=0.08 m, as shown by the dashed line in the top inset image. Without the scheme of enhancing the mutual diffusivity proposed in [38], the gas component in gas phase would exhibit non-uniform distributions with discontinuous bands even after a long simulation time. With the modified interaction forces in Eqs. 14-15, it can be observed in Fig. 3 that the density field is quite uniform and no discontinuous bands can be observed in the gas phase. From the density profile it can be seen that the liquid phase is mainly occupied by the liquid component and the gas phase is mainly occupied by the gas component. The density ratio between the liquid phase and gas phase is about 152.1 (8.17/0.0537).

**Fig. 3** Density of liquid and gas components along the y axis. The density is the averaged density of the densities at $x$=0.02 and $x$=0.08 m, as shown by the dashed line in the top inset image. The top and bottom inset images are the density field of the liquid component and that of the gas component, respectively. The circle located at (0.05m, 0.05m) is the $D_{(s)}$ particle

*5.2 Effects of different reactions*

*5.2.1. Decomposition of* $A_{(aq)}$

We then simulate the reactive transport processes with only R1 turned on in the system. The two-phase system established in Section 5.1 is used as the input of the simulation. Figs. 4a-4c shows the time evolutions of the concentration contours for $A_{(aq)}$, $B_{(aq)}$ and $C_{(g)}$. As can be seen from the figure, $A_{(aq)}$ is decomposed more quickly near the $D_{(s)}$ surface due to higher decomposition rate there, leading to higher concentration of $B_{(aq)}$ and $C_{(g)}$ around $D_{(s)}$. It is clearly from Fig. 4(a-b) that the liquid-gas interface serves as a non-flux boundary for the transport of $A_{(aq)}$ and $B_{(aq)}$. In Fig. 4(c), due to the lack of nucleation mechanism for gas bubble in the present model, nucleation and growth of $C_{(g)}$ bubble in the bulk liquid is not captured. It is expected that if the bubble emerges, parts of the $D_{(s)}$ surface will be covered which will not be available for decomposition. The generated $C_{(g)}$ transports from the liquid phase, crosses the liquid-gas interface and finally arrives at the gas phase. Concentration jump of $C_{(g)}$ can be clearly observed across the liquid-gas interface, indicating that the Henry's law for $C_{(g)}$ is obeyed in the current model. Because $C_{(g)}$ generated above the $D_{(s)}$ particle can transport more freely into the gas phase than that below the $D_{(s)}$ particle, thus the concentration of $C_{(g)}$ above the particle is lower than that below the particle, as shown in Fig. 4(c). Such distributions lead to quicker decomposition rate in the upper hemisphere of $D_{(s)}$ than in the lower hemisphere according to the reaction kinetics of R1. Finally, the mass of liquid phase continuously declines due to the loss of mass of $C_{(g)}$, resulting in a fall of the liquid-gas interface from $y = 100$ to $y = 90$, as can be observed in Fig. 4.

Fig. 4 Time evolutions of concentration fields of (a) $A_{(aq)}$, (b) $B_{(aq)}$ and (c) $C_{(g)}$ with only decomposition reaction R1 considered

*5.2.2. Effects of dissolution reaction*

We now examine the effects of dissolution reaction R2 further turned on. Again the two-phase system established in Section 5.1 is used as the input. Figs. 5(a)-5(d) display the time evolutions of the concentration contours of $A_{(aq)}$, $B_{(aq)}$, $C_{(g)}$ and $E_{(aq)}$, respectively. Compared with Fig. 4, the notable phenomenon is the shrink of the $D_{(s)}$ particle due to dissolution reaction. The surface dissolution reaction and the shrink of $D_{(s)}$ has a significant effect on the surface area of $D_{(s)}$ and thus on the reactions. First, due to the consumption of $B_{(aq)}$ around the $D_{(s)}$ particle, its concentration near the particle is first higher and then becomes lower than that far away from the particle, as shown in Fig. 5(b). Second, the dissolution reaction results in the generation of $E_{(aq)}$, which gradually increases around the $D_{(s)}$ particle, as shown in Fig. 5(d). Finally, once the solid particle is completely dissolved, the decomposition rate in the entire liquid phase is the same, and thus concentration fields of all the components approach uniform distribution, as shown at $t=700000$.

**Fig. 5** Time evolutions of concentration fields of (a) $A_{(aq)}$, (b) $B_{(aq)}$, (c) $C_{(g)}$ and (d) $E_{(aq)}$ with decomposition reaction R1 and dissolution reaction R2 considered

*5.2.3. Effects of precipitation reaction*

We further take into account the precipitation reactions R3 and R4 (Now all the reactions are considered). Again the two-phase system established in Section 5.1 is used as the input. Figs. 6(a)-6(e) display the time evolutions of the concentration contours of $A_{(aq)}$, $B_{(aq)}$, $C_{(g)}$ and $E_{(aq)}$ and $F_{(aq)}$, respectively. The circle dot around the surface of $D_{(s)}$ in Fig. 6(f) denotes the precipitates. Due to reaction R3, it can be seen that $F_{(aq)}$ is generated around the solid particle $D_{(s)}$, as shown in Fig. 6(e). The peak region of the concentration of $F_{(aq)}$ is firstly located near the $D_{(s)}$ particle as concentration of $B_{(aq)}$ and $E_{(aq)}$ is higher there ($t=4000$, Fig. 6(e)), then it departs from the $D_{(s)}$ particle due to the Reaction R4 which consumes $F_{(aq)}$ and generates $F_{(s)}$ ($t=10000$, Fig. 6(e)). At the initial stage, the precipitated solid $F_{(s)}$ is sporadic and thus has limited effects on the dissolution reaction on the surface of $D_{(s)}$ ($t=10000$, Fig. 6(f)). As more $F_{(aq)}$ is produced, more $F_{(s)}$ on the surface of $D_{(s)}$ is generated. Therefore, $F_{(s)}$ gradually becomes compact ($t=20000$, Fig. 6(f)) and finally covers the entire surface of $D_{(s)}$ ($t=100000$, Fig. 6(f)). Such coverage leads to the following phenomena. First, the surface of $D_{(s)}$ is completely separated from the bulk liquid. Thus, the faster decomposition rate of $A_{(aq)}$ around the $D_{(s)}$ is no longer available and the decomposition rate is uniform in the entire liquid phase, leading to relatively uniform distribution of $A_{(aq)}$ ($t=100000$, Fig. 6(a)). Second, the dissolution reaction R2 is halted and thus $E_{(aq)}$ is no longer produced. The concentration of $E_{(aq)}$ gradually decreases as it is still being consumed according to Reaction R3 ($t=20000$ and $t=100000$, Fig. 6(d)). Third, when $E_{(aq)}$ is completely consumed, reaction R3 stops, leading to only Reactions R1 and R4 existing in the system. As a result, the concentration of $B_{(aq)}$ gradually increases ($t$=100000 and $t$=700000, Fig.6(b)), exhibiting uniform distributions ultimately. Meanwhile, $F_{(aq)}$ is completely converted to $F_{(s)}$ ($t$=700000, Fig. 6(e)), leading to rich precipitation patterns.

The precipitation pattern can be described as compact coral-type structures at the initial stage ($t$=20000 and $t$=100000, Fig. 6(f)) and then as open cluster-type structures ($t$=700000, Fig. 6(f)). Such patterns indicate that concentration of reactants required for precipitation reaction is relatively sufficient at first and becomes deficient later on, and the reactive transport process transforms from a reaction-controlled process to a diffusion-controlled process [14]. Indeed, $F_{(aq)}$ is firstly sufficient when the $D_{(s)}$ surface is not entirely covered by $F_{(s)}$ ($t$=4000~20000, Fig. 6(e)) and then becomes insufficient due to the cease of Reactions R2 and R3 ($t$=20000~700000, Fig. 6(e)). Another interesting phenomenon is that the precipitation also leads to the trap of the components ($t$=100000 and $t$=700000). It is worth pointing out that in the trapped fluid, the concentration of $B_{(aq)}$ is relatively low and that of $E_{(aq)}$ is relatively high as shown at $t$=100000, due to the dissolution reaction R2. Thus, $E_{(aq)}$ there cannot be completely reacted according to Reaction R3; but will remain there even at the end of the simulations ($t$=700000, Fig. 6(d)). The trapped $C_{(g)}$ cannot escape either. Finally, it can be observed that the liquid-gas interface rises in Fig. 6 rather than drops in Figs. 4-5. This is because the precipitates take the place once occupied by the liquid phase, proving the effectiveness of the scheme for treating the information of the nodes undergoing phase change introduced in Section 4.5.

**Fig. 6** Time evolutions of concentration fields of (a) $A_{(aq)}$, (b) $B_{(aq)}$, (c) $C_{(g)}$, (d) $E_{(aq)}$ and (e) $F_{(aq)}$ and of (f) the geometries of $D_{(s)}$ and $F_{(s)}$ with all the reactions taken into account. The green node in Fig. 6(f) is $D_{(s)}$ and the black node denotes $F_{(s)}$

*5.3 Evolutions of concentration and solid volume*

Discussion in Section 5.2 presents a phenomenological description of the physicochemical problem. Fig. 7 quantitatively shows the time evolutions of the total concentrations of different components. In Fig. 7, the suffix 1, 2 and 3 denotes reactions with only decomposition (corresponding to Section 5.2.1), with dissolution added (corresponding to Section 5.2.2), and with precipitation further added (corresponding to Section 5.2.3), respectively. For only decomposition reaction R1 considered, it is expected that the curves of A_1 and B_1 show the symmetrical characteristics, as the amount of decomposed A_1 is equal to that of generated B_1. The concentrations of A_1 and B_1 at the intersection point is about $1.1\times10^8$ mol m$^{-3}$, just half of the initial concentration of A_1 (about $2.2\times10^8$ mol m$^{-3}$), further validating our numerical method.

**Fig. 7** Time evolutions of the total concentrations of $A_{(aq)}$, $B_{(aq)}$, $E_{(aq)}$ and $F_{(aq)}$

When dissolution reaction R2 is turned on, it can be seen that the slope of A_2 is first greater and then smaller than that of A_1, indicating the decomposition rate of A_2 is first faster and then slower than that of A_1. This is due to that the dissolution reaction has two opposite effects on the decomposition reaction. On one hand, the dissolution reaction consumes $B_{(aq)}$, thus accelerating Reaction R1. On the other hand, the surface area of $D_{(s)}$ decreases as the dissolution progresses, which reduces the decomposition rate. Compared A_2 with A_1, it can be concluded that initially the favorable effect dominates and then the adverse effect does. After t~200000, the slope of A_2 becomes constant, so does E_2 though it initially increases with time. This point of time corresponds to the complete dissolution of $D_{(s)}$, in good agreement with Fig. 6. After this point reaction R2 stops due to the lack of $D_{(s)}$, and reaction R1 only takes place in the bulk liquid.

Finally, when the precipitation reactions R3 and R4 are further taken into account (suffix 3), the favorable effect of the dissolution reaction (accelerating Reaction R1 towards the right side) is observed at the initial stage. However, in a short period of time the precipitates cover the surface of $D_{(s)}$ and separate it from the fluid, thus slowing down the decomposition rate. Due to the coupling mechanism between precipitation and dissolution reactions, E_3 and F_3 increase to a peak value and then decrease to almost zero. Unlike B_2 which increases during the entire period of time, B_3 remains almost zero initially because it is consumed by both R2 and R3 reactions. After the surface of $D_{(s)}$ is fully covered by $F_{(s)}$ and $E_{(aq)}$ is completely consumed subsequently (*t*~200000), both R2 and R3 which consume $B_{(aq)}$ stop, and thus concentration of $B_{(aq)}$ gradually rises.

Fig. 8 shows the time evolutions of the volume of $D_{(s)}$ and $F_{(s)}$. Compared with D_1 and D_3, the dissolution of D_2 is the quickest and the most thorough, which is expected based on the above discussion. The final volume of F_3 is several times higher than that of D_3, due to the higher molar volume of F_3 and the extremely low saturation concentration of $F_{(aq)}$.

**Fig. 8** Time evolutions of the total volume of $D_{(s)}$ and $F_{(s)}$

*5.4 Effects of precipitation reaction rate constant*

We further investigate the effects of the precipitation reaction rate constant. Fig. 9 shows the time evolutions of the geometries of $D_{(s)}$ and $F_{(s)}$ for $k_4/10000$ and $k_4/20000$, respectively, where $k_4$ is the reaction rate constant of R4. For simplicity, the time variations of the concentration fields are not shown here. It can be seen that the precipitation pattern in Fig. 9(a) is more like compact-annulus structures around the un-dissolved $D_{(s)}$, rather than the open cluster-type structures shown in Fig. 6(e). This means that the reactive transport processes now are reaction-controlled, which is expected as the precipitation reaction rate constant is reduced by ten thousand times. Besides, it can be seen that the precipitates are relatively porous initially, and liquid components can transport through the porous layer of precipitates to arrive at the underlying surface of $D_{(s)}$. As a result, the dissolution of $D_{(s)}$ is still allowed ($t$=100000~200000). However, as the reaction progresses, the precipitates become more and more compact and finally completely cover the surface of $D_{(s)}$ and stop the dissolution ($t$=700000~900000). By further reducing the reaction rate constant of R4 to $k_4/20000$, it can be seen in Fig. 9(b) that the generation rate of $F_{(s)}$ is further decreased and this time the precipitates are porous enough throughout the entire simulation time, and $D_{(s)}$ can be completely dissolved.

**Fig. 9** Time evolutions of the geometries of $D_{(s)}$ and $F_{(s)}$ for (a) $k_4/10000$ and (b) $k_4/20000$, respectively, where $k_4$ is the reaction rate of R4 in Section 5.2.3

Fig. 10 shows the time variations of the total solid volume of $D_{(s)}$ and $F_{(s)}$ for different precipitation reaction rate constants. In the figure, suffix _3 denotes the rate used in Section 5.2.3, _4 is that for $k_4/10000$ and _5 represents that for $k_4/20000$. It can be seen that decreasing the precipitation reaction rate constant leads to an increase in the dissolution rate and in the total amount of dissolved $D_{(s)}$. This is expected because fewer surfaces are covered by the precipitates at a smaller precipitation rate constant. However, for precipitation of $F_{(s)}$, although the rate decreases as the precipitation reaction rate constant is reduced, the total amount of precipitation increases. This again demonstrates the coupling characteristics of the precipitation and dissolution reactions.

**Fig. 10** Time variations of the total solid volume of $D_{(s)}$ and $F_{(s)}$ for different precipitation reaction rate. Suffix _3 denotes that in Section 5.3, _4 is that for $k_4/10000$ and _5 represents that for $k_4/20000$

# 6. Conclusion

A two-phase multi-mixture model based on the LBM is developed for multicomponent multiphase reactive transport with dissolution-precipitation processes that are widespread in energy and environment science. The model combines the pseudopotential LB multiphase model with large density ratio, the mass transport LB model and a front-tracking method called VOP for updating solid structures. In this model, instead of directly including every component in the construction of the two-phase system, each phase (liquid or gas) is considered as a mixture of the corresponding components which are miscible in it. In other words, the two-phase system is constructed using only a unified liquid component and a unified gas component. This way, the tedious work of tuning a considerable number of parameters is avoided when all the components are included to construct the multiphase system based on the pseudopotential LB multiphase model. After the two-phase system is established, the mass transport of each component in the corresponding phase is solved using the mass transport LB model. The dynamic evolutions of the liquid-solid interfaces due to dissolution-precipitation reactions are captured using the VOP.

The two-phase multi-mixture model is adopted to simulate a physicochemical problem which contains liquid-gas-solid three phases including one gas component, four aqueous components, two solid phases, and four reactions including two bulk reactions, a dissolution reaction and a precipitation reaction. Pore–scale multiphase fluid flow, multicomponent mass transport, homogeneous bulk reactions and heterogeneous reactions, and the dissolution-precipitation of the solid phases are well captured by the pore-scale two-phase multi-mixture model. Effects of the decomposition, dissolution and precipitation reactions on the multicomponent multiphase reactive transport processes are investigated and discussed in detail. The simulation results demonstrate the capacity of the two-phase multi-mixture model, and reveal complex and strong interactions among different sub-processes.

## Acknowledgements

Financial support of this work was provided by Shell International Exploration & Production, Inc., the LANL LDRD Program, the UC Lab Fees Research Program, the National Nature Science Foundation of China (No. 51406145 and No. 51136004) and the NNSFC international-joint key project (No. 51320105004). We also acknowledge the support of LANL's Institutional Computing Program.

## Appendix A

*1. Validation of the multicomponent multiphase pseudopotential model*

Not until recently has the pseudopotential model been applied to multicomponent multiphase fluid flow problems with large density ratios [33, 34, 37]. The model was not tested against the Laplace law in any of these studies. In addition, only Bao and Schaefer [34] applied the model to simulate contact angle problems, while the solid-fluid interaction was absent in other studies [33, 37]. In the present study, two representative problems, namely a single bubble surrounded by liquid in a fully periodic domain and static contact angle on a horizontal homogenous solid surface, are simulated to validate the multicomponent multiphase pseudopotential model with large density ratios.

First, the Laplace law is tested by placing a bubble with different radiuses in a fully periodic domain filled with liquid. Initially, the density of the liquid component is set as 8.16 and 0.05 outside and inside the bubble, respectively, while that for gas component is set as 0.0001 and 0.05 outside and inside the bubble, respectively. After steady state is obtained, the pressure is calculated by

$$p = c_s^2 \sum \rho_\sigma + \frac{1}{2} c_s^2 \sum G_{\sigma\sigma} \psi_\sigma^2 + \frac{1}{2} c_s^2 \sum_{\bar\sigma \neq \sigma} G_{\sigma\bar\sigma} \varphi_\sigma \varphi_{\bar\sigma} \qquad (A1)$$

Fig. A1(a) shows a typical density distribution field of the gas component and the velocity vectors obtained from the simulation. The spurious currents near the phase interface can be clearly seen, which is mainly in the gas phase. The maximum spurious current in lattice unit is 0.05. Note that for the relatively flat interface shown in Fig. 4, as the curvature approaches zero, the maximum spurious current is extremely low, with the magnitude of $10^{-10}$ in lattice unit. Fig. A1 (b) shows the pressure distribution along $x$ direction at $y=H/2$, where $H$ is the height of the domain. Due to the large density variation near the phase interface, pressure fluctuates sharply, leading to unphysical value, which is also observed in the simulation using single component multiphase pseudopotential model with large density ratios [4]. Such an unphysical fluctuation is not accounted for in the calculation of the pressure difference. Fig. A1(c) plots the pressure difference $\Delta p$ as a function of $1/r$ with $r$ as the radius of the bubble. The linear relationship can be clearly seen and the slope is 0.667, which is the surface tension coefficient $\sigma$ according to Laplace law $\Delta p = \sigma/r$.

**Fig. A1** Simulation results of a bubble in a fully periodic domain filled by liquid component. (a) Density distribution field and the velocity vectors. (b) Pressure distribution along the horizontal center of the computational domain. (c) Dependence of capillary pressure on bubble radius and calibration of the Laplace law

Second, equilibrium contact angle of liquid on a homogeneous horizontal solid wall is simulated. Contact angle is usually considered as a measure of the solid surface wettability. A surface is wetting or hydrophilic if the contact angle $\theta<90°$, and liquid tends to spread as a film on the solid surface. In contrast, the surface is non-wetting or hydrophobic if $\theta>90°$, and liquid tends to form a droplet on the solid surface. A series of simulations are carried out in which an initially semicircular static droplet is placed on a horizontal solid surface and the solid-fluid interaction, $w$ is changed in the range of $-0.4$~$0.1$ to obtain different contact angles. The simulations are performed in a 201×201 lattice system with the top and bottom boundaries as solid walls and the left and right boundaries as periodic boundaries. Fig. A2(a) presents the liquid droplets with different contact angles by changing $w$. Fig. A2 (b) shows the relationship between *w* and the predicted contact angle. The functional relationship is almost linear, which is also observed in simulations using multicomponent multiphase pseduopotetial model with equal density ratio or using single-component multiphase pseudopotential model with large density ratio. Note that in the present study the density ratio between liquid component and gas component is about 150. We found that with such a large density ratio the range of the contact angle that can be achieved in our simulation is limited to (70°, 100°). The limited contact angle range under large density ratio is also found in [34]. Further work needs to be done to overcome this limitation for practical applications.

**Fig. A2** Effects of solid-fluid interaction *w* on contact angle: (a) Liquid droplet with different contact angles: left: w=-0.4, θ=74.2°, middle: w=0, θ=90°; right: w=0.14, θ=97°; (b) the relationship between contact angle and solid-fluid interaction *w*

*2. Validation of the LB mass transport model*

The LB mass transport model used in the present study has been well validated in several of our previous studies. For the purpose of brevity, the validation is not repeated here and one can refer to [4, 36, 43] for more details.

*3. Validation of the dissolution/precipitation model*

Reactive transport processes including the dissolution of a single mineral into a solution is simulated to validate our code for tracking solid-liquid interface. The simulation geometry is a fractured medium with size of *L\*H*. The initial fracture aperture is $h_0$. For fluid flow, at the entrance (*x*=0) and exit (*x*=100), the pressure is specified; at *y*=0 and *y*=90, no slip boundary condition is adopted. Initially, the solution is saturated and the solid does not dissolve. When fluid flow reaches a steady state, the inflowing fluid changes to a pure solvent, and then the solid begins to dissolve. The reaction at the solid-fluid interface is the first-order kinetic reaction as follows

$$D\frac{\partial C}{\partial n} = k(C - C_s) \tag{A2}$$

where $C$ is the solute concentration at the interface and $C_s$ is the saturated saturation. $k$ is the reaction rate constant and $D$ is the diffusivity.

Fluid flow between two parallel static walls is called Poiseuille flow and an analytical solution exists for the permeability $k_h$, namely $k_h = h^2/12$, where $h$ is the distance between the two walls or is the fracture aperture in the present study. Therefore, the permeability of the domain simulated is $k_H = (h^2/12) \cdot (h/H)$. $k_H$ normalized by its initial value $k_H^0$ can be written as

$$\frac{k_H}{k_H^0} = \left(\frac{h/H}{h_0/H}\right)^3 = \left(\frac{\varepsilon}{\varepsilon_0}\right)^3 \tag{A3}$$

where $\varepsilon = h/H$ is the porosity of the fractured medium. Eq. (A3) indicates that the normalized permeability and porosity of the fractured medium obey a power law with an exponent of 3. We try to validate our code based on Eq. (A3). In the simulation, $L$=100, $H$=90 and $h_0$=30 lattice units. Note that Eq. (A3) only applies when $L$ remains constant and $h$ does not change along its length. In other words, dissolution only occurs at the channel walls uniformly but not on the vertical surfaces at the inlet and outlet boundaries. In the simulation, we set these two vertical surfaces unreactive, and fix the $Da$ at a very small value (0.001) to ensure uniform dissolution of the fracture walls. Fig. A3 shows the normalized permeability and porosity relationship from both LBM simulation (square) and from Eq. (A3) (solid line). Clearly the LBM result is in good agreement with the analytical solution.

**Fig. A3** Relationship between dimensionless porosity and permeability with extremely low $Da$

# Figure and Table Caption

Fig. 1 (Color online) Schematic of the multicomponent multiphase reactive transport processes

Fig. 2 (Color online) Dynamic evolutions of liquid-gas-solid interfaces during the multicomponent multiphase reactive transport processes. There are three types of nodes in the domain: solid nodes, liquid nodes, and gas nodes; and there are four kinds of changes of node type: namely liquid to solid due to precipitation, solid to liquid due to dissolution, liquid to gas, and gas to liquid. The last two of these changes of node type are due to evolution of the liquid-gas interfaces. Note that there is no exchange between gas node and solid node as dissolution and precipitation only take place at liquid-solid interface.

Fig. 3 (Color online) Density of liquid and gas components along the y axis. The density is the averaged density of the densities at *x*=0.02 and *x* =0.08 m, as shown by the dashed line in the top inset image. The top and bottom inset images are the density field of the liquid component and that of the gas component, respectively.  The circle located at (0.05m, 0.05m) is the $D_{(s)}$ particle.

Fig. 4 (Color online) Time evolutions of concentration fields of (a) $A_{(aq)}$, (b) $B_{(aq)}$ and (c) $C_{(g)}$ with only decomposition reaction R1 considered.

Fig. 5 (Color online) Time evolutions of concentration fields of (a) $A_{(aq)}$, (b) $B_{(aq)}$, (c) $C_{(g)}$ and (d) $E_{(aq)}$ with decomposition reaction R1 and dissolution reaction R2 considered.

Fig. 6 (Color online) Time evolutions of concentration fields of (a) $A_{(aq)}$, (b) $B_{(aq)}$, (c) $C_{(g)}$, (d) $E_{(aq)}$ and (e) $F_{(aq)}$ and of (f) the geometries of $D_{(s)}$ and $F_{(s)}$ with all the reactions taken into account. The green node in Fig. 6(f) is $D_{(s)}$ and the black node denotes $F_{(s)}$.

Fig. 7 (Color online) Time evolutions of the total concentrations of $A_{(aq)}$, $B_{(aq)}$, $E_{(aq)}$ and $F_{(aq)}$.

Fig. 8 (Color online) Time evolutions of the total volume of $D_{(s)}$ and $F_{(s)}$.

Fig. 9 (Color online) Time evolutions of the geometries of $D_{(s)}$ and $F_{(s)}$ for (a) $k_4/10000$ and (b) $k_4/20000$, respectively, where $k_4$ is the reaction rate of R4 in Section 5.3.

Fig. 10 (Color online) Time variations of the total solid volume of $D_{(s)}$ and $F_{(s)}$ for different precipitation reaction rate. Suffix _3 denotes that in Section 5.3, _4 is that for $k_4/10000$ and _5 represents that for $k_4/20000$.

Fig. A1 Simulation results of a bubble in a fully periodic domain filled by liquid component. (a) Density distribution field and the velocity vectors. (b) Pressure distribution along the horizontal center of the computational domain. (c) Dependence of capillary pressure on bubble radius and calibration of the Laplace law.

Fig. A2 Effects of solid-fluid interaction $w$ on contact angle: (a) Liquid droplet with different contact angles: left: w=-0.4, $\theta=74.2°$, middle: w=0, $\theta=90°$; right: w=0.14, $\theta=97°$; (b) the relationship between contact angle and solid-fluid interaction w

Fig. A3 Relationship between dimensionless porosity and permeability with extremely low $Da$

Table 1: Parameters used in the simulations

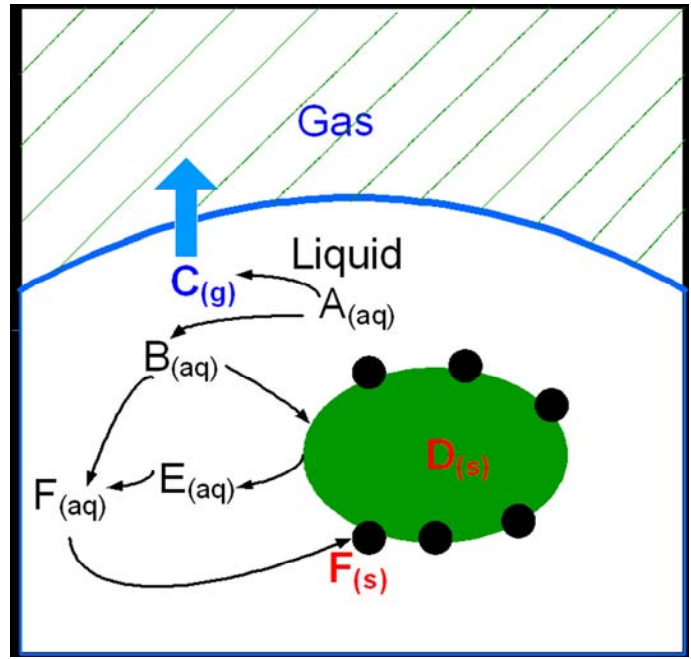

Fig. 1

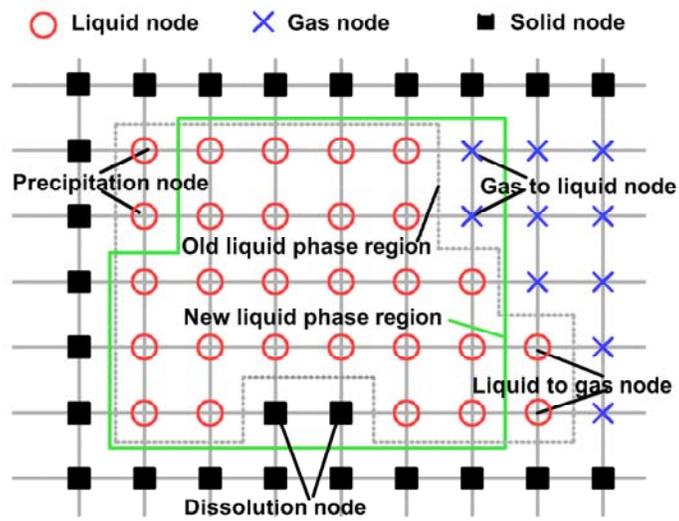

Fig. 2

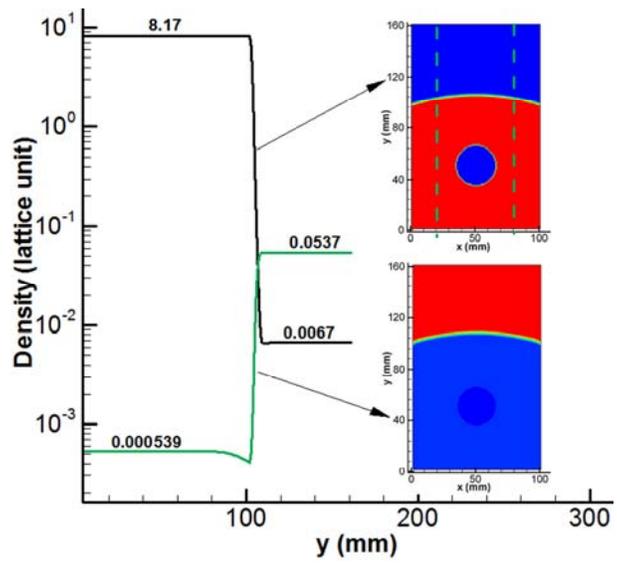

Fig. 3

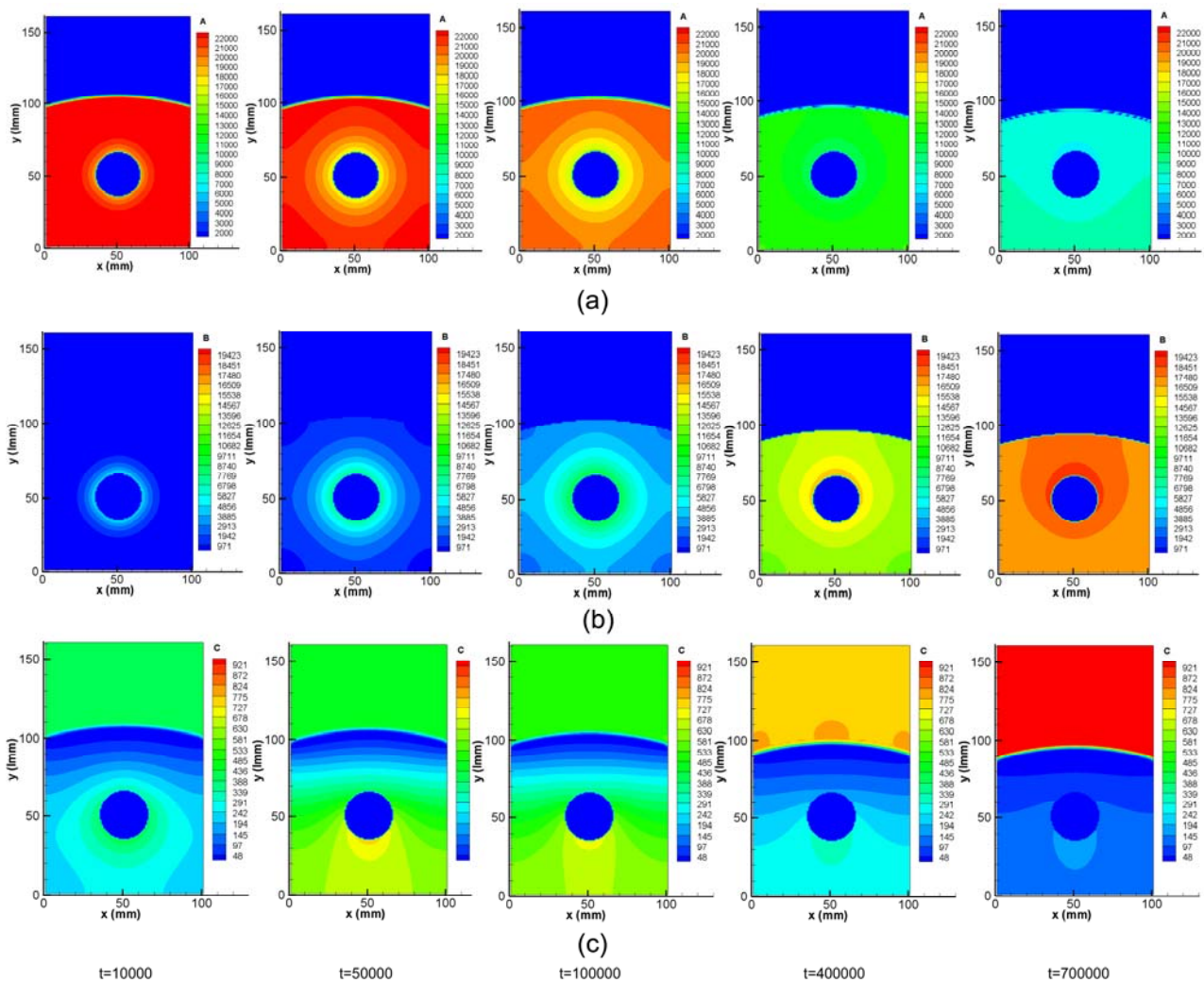

Fig. 4

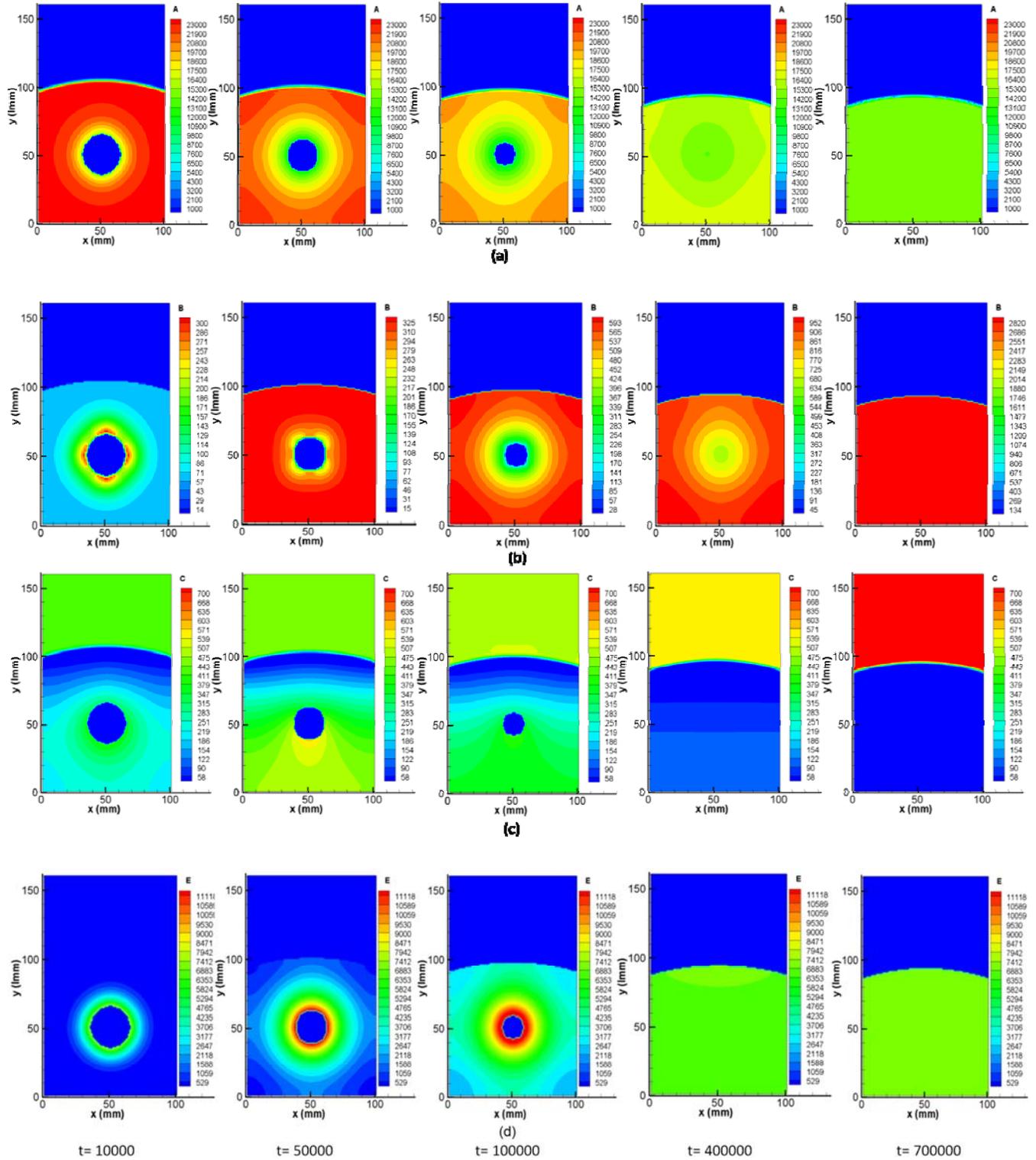

Fig. 5

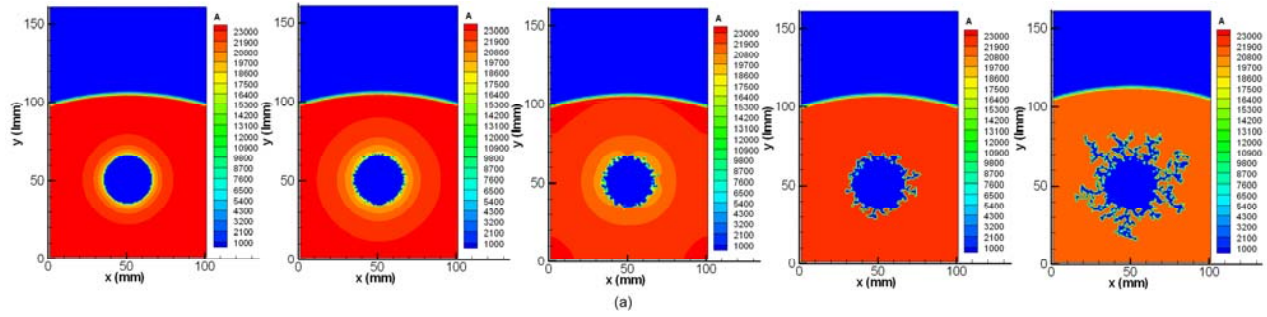

(a)

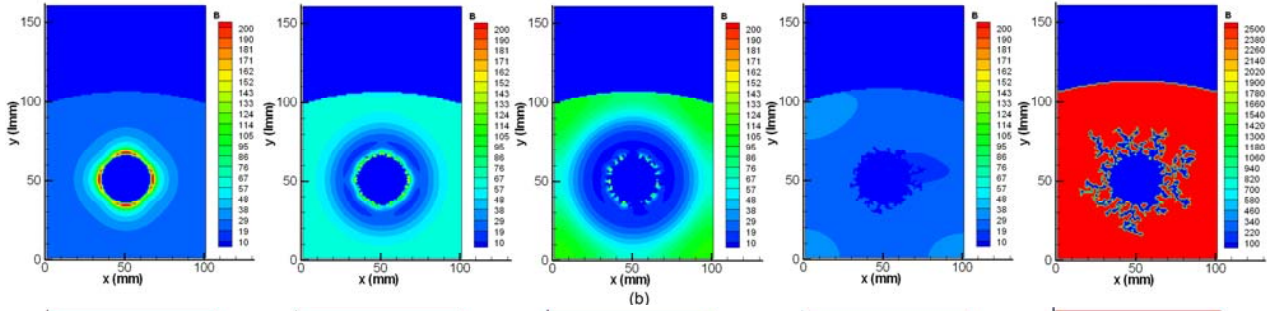

(b)

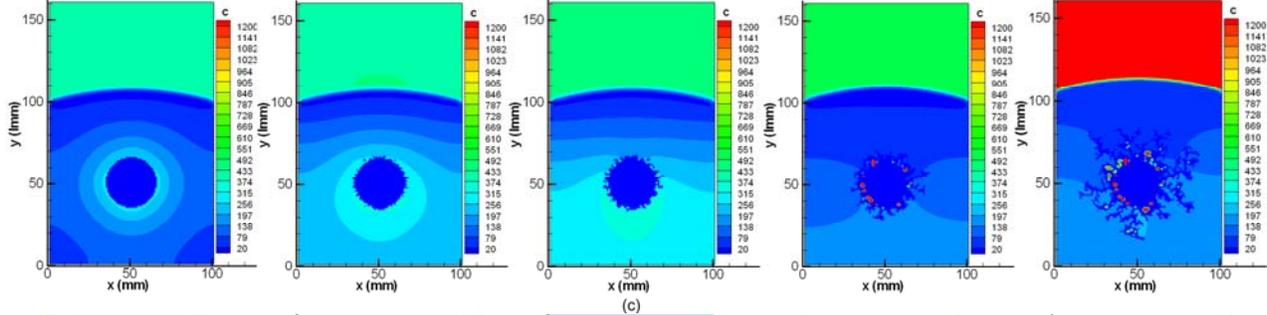

(c)

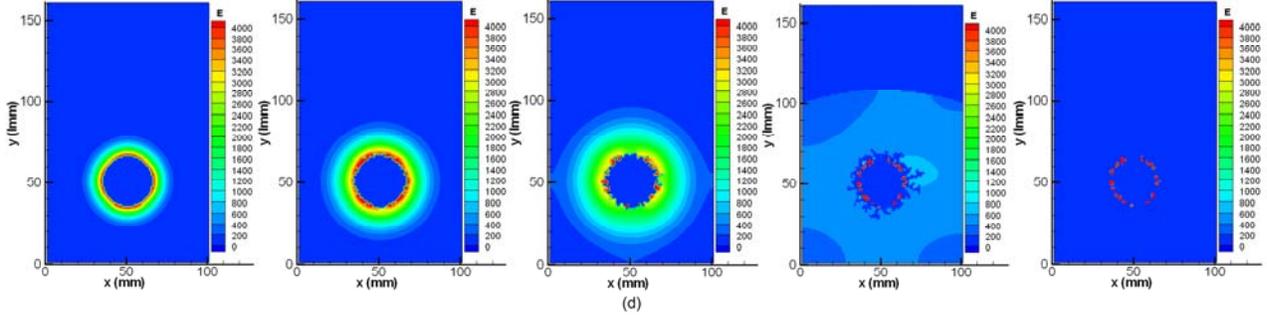

(d)

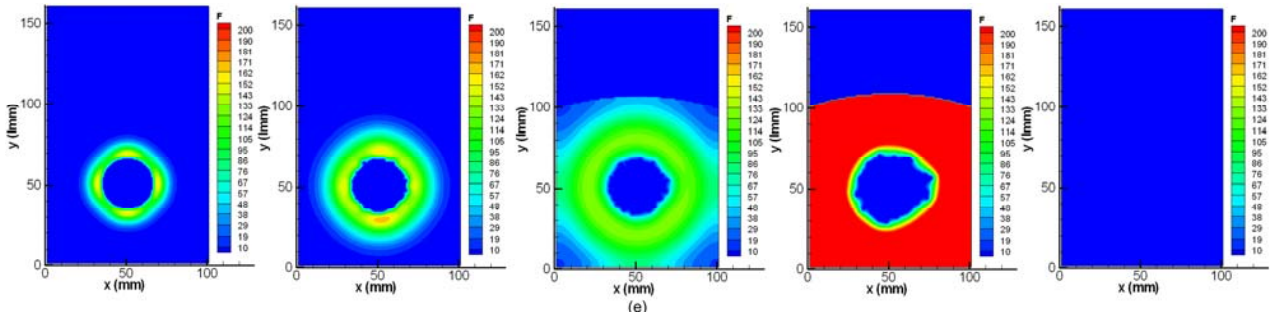

(e)

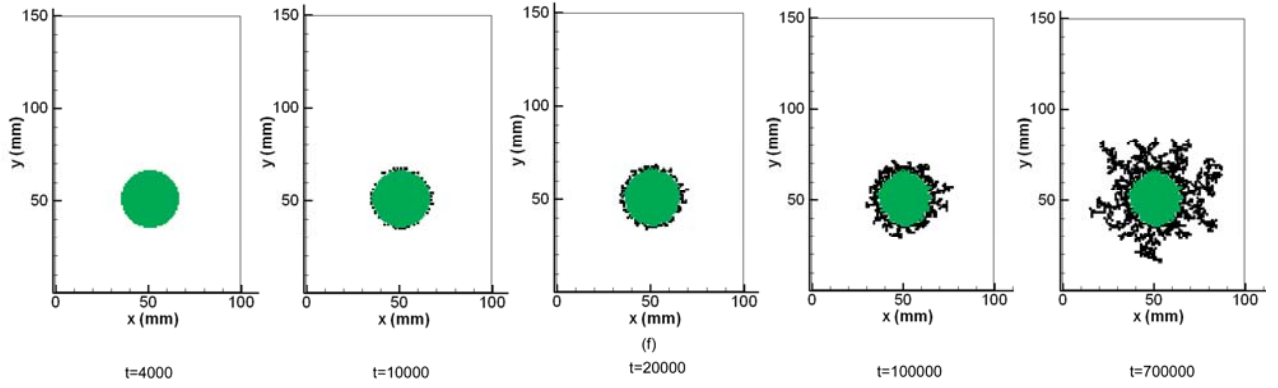

Fig. 6

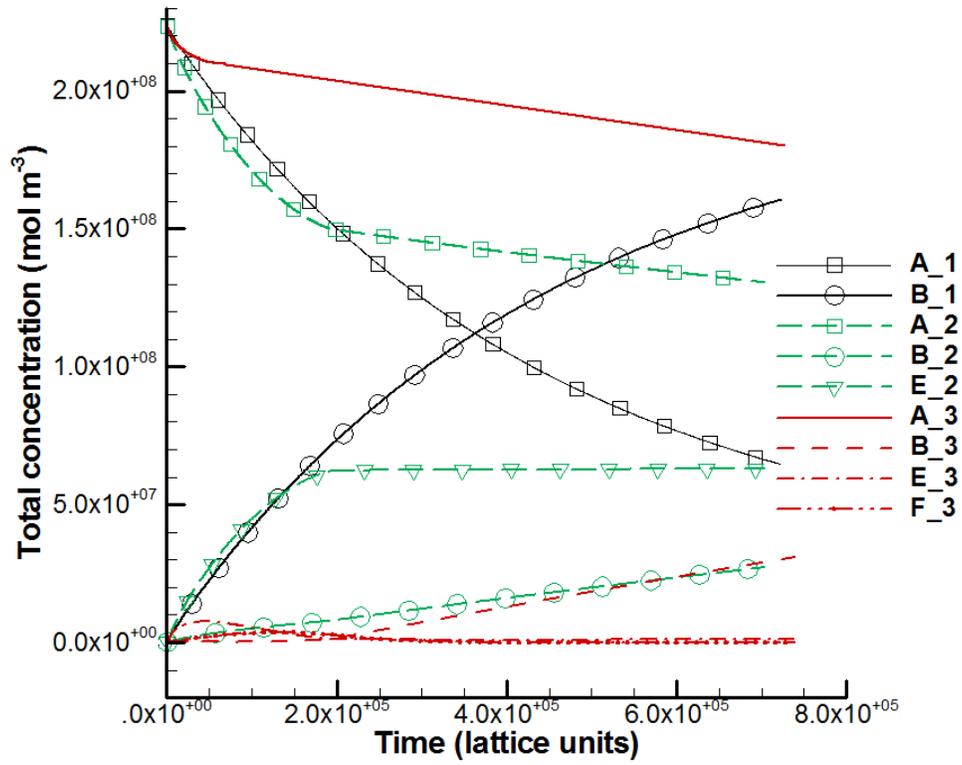

Fig. 7

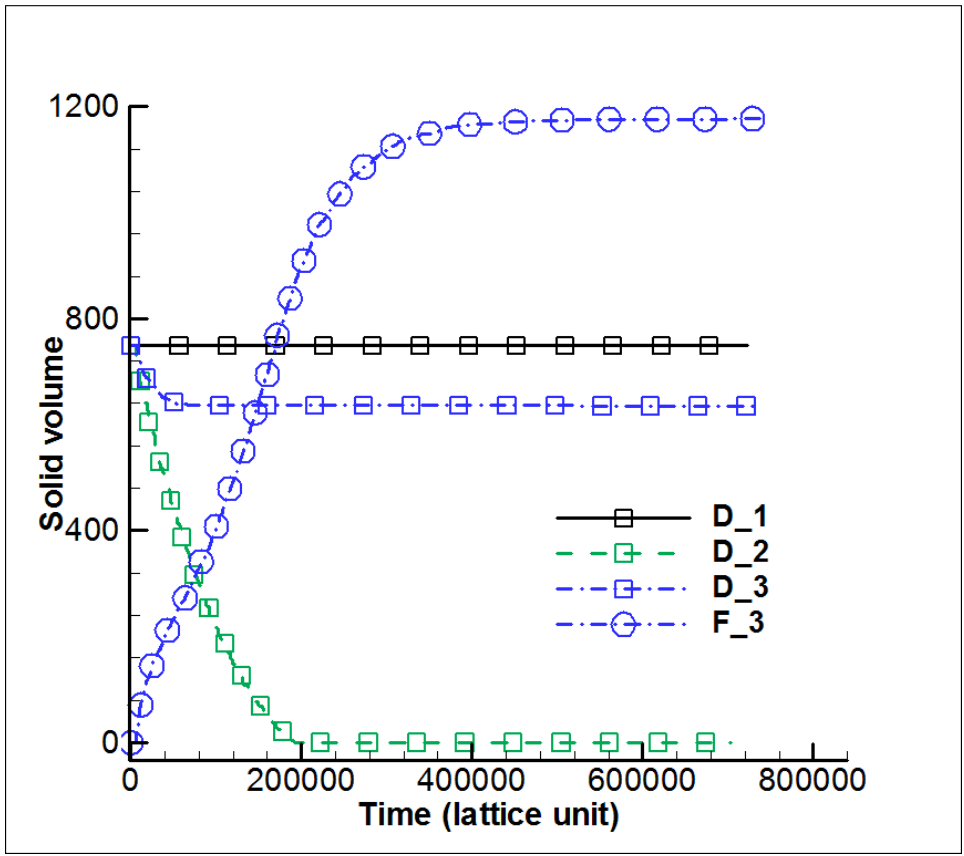

Fig. 8

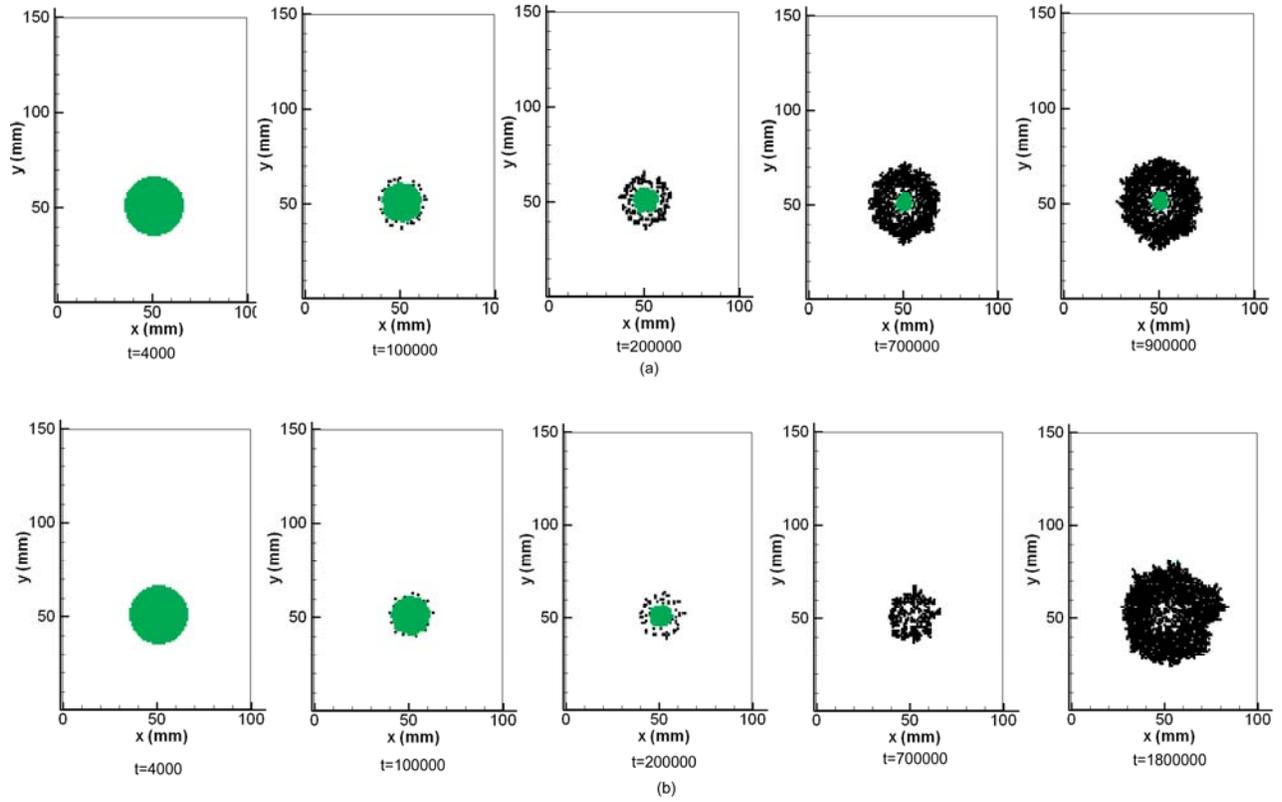

Fig. 9

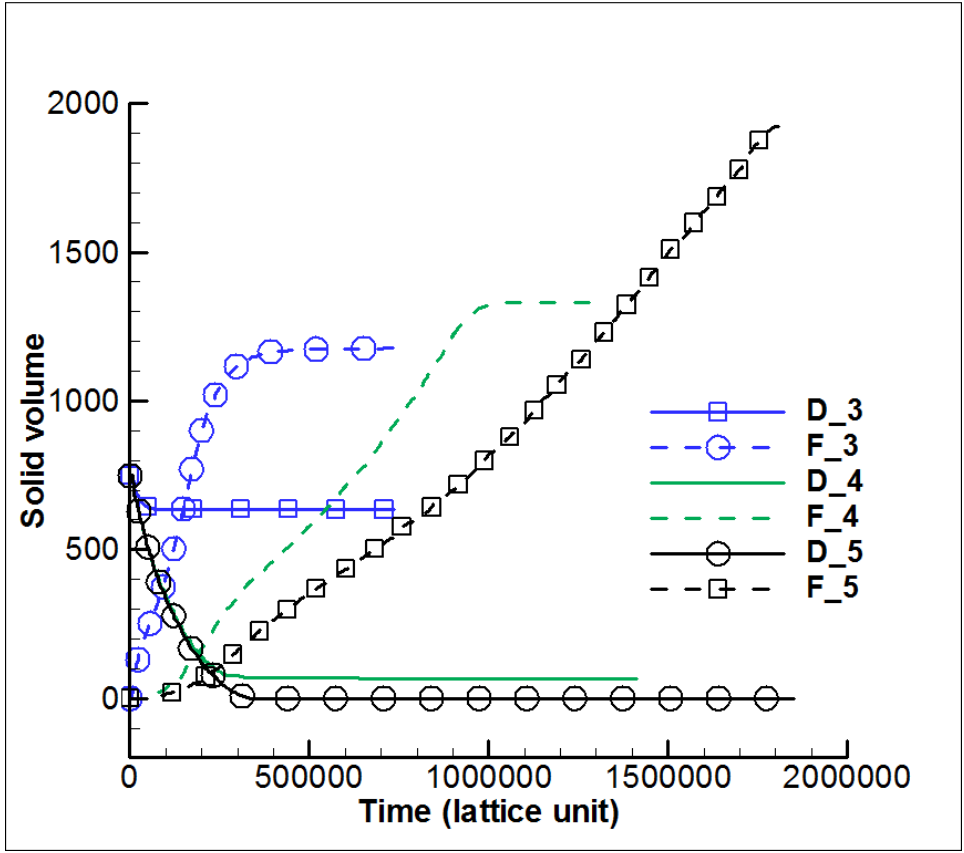

Fig 10

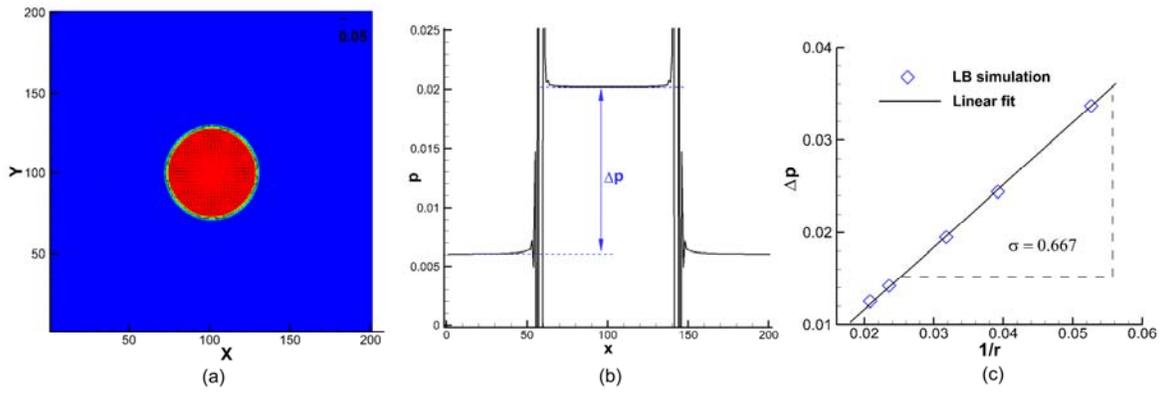

Fig. A1

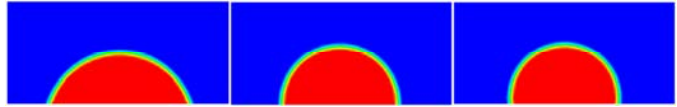
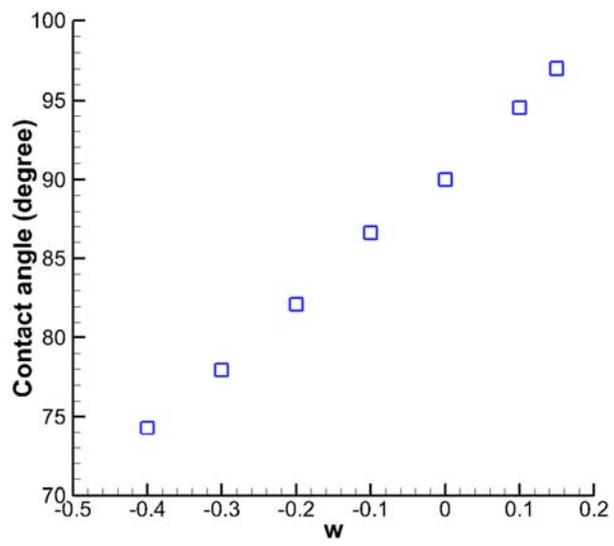

Fig. A2

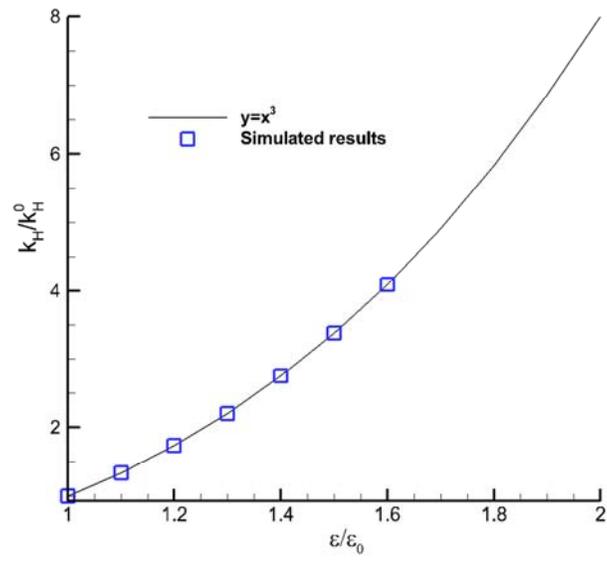

Fig. A3